\documentclass[conference]{IEEEtran}
\IEEEoverridecommandlockouts
\usepackage[hyphens]{url}
\usepackage{hyperref}
\hypersetup{breaklinks=true}
\usepackage{tcolorbox} 
\usepackage{mathtools}

\usepackage{tikz}
\usepackage{textcomp}
\usepackage{hyperref}

\renewcommand{\figurename}{Figure}

\newcommand\copyrighttext{%
  \footnotesize \textcopyright 2017 IEEE. Personal use of this material is permitted.
  Permission from IEEE must be obtained for all other uses, in any current or future 
  media, including reprinting/republishing this material for advertising or promotional 
  purposes, creating new collective works, for resale or redistribution to servers or 
  lists, or reuse of any copyrighted component of this work in other works.}
\newcommand\copyrightnotice{%
\begin{tikzpicture}[remember picture,overlay]
\node[anchor=south,yshift=20pt] at (current page.south) {\fbox{\parbox{\dimexpr\textwidth-\fboxsep-\fboxrule\relax}{\copyrighttext}}};
\end{tikzpicture}%
}

\usepackage{graphicx}

\ifCLASSINFOpdf
\else
\fi
\begin{document}

\title{Why the Equifax Breach \\ Should Not Have Mattered}

\author{\IEEEauthorblockN{Marten Lohstroh\thanks{This work was supported in part by the TerraSwarm Research Center, one of six centers supported by the STARnet phase of the Focus Center Research Program (FCRP) a Semiconductor Research Corporation program sponsored by MARCO and DARPA.}} \\
\IEEEauthorblockA{Department of Electrical Engineering and Computer Science\\
University of California, Berkeley\\
Email: marten@berkeley.edu}
}


%


\maketitle
\copyrightnotice
\begin{abstract}
Data security, which is concerned with the prevention of unauthorized 
access to computers, databases, and websites, helps protect digital 
privacy and ensure data integrity. It is extremely difficult, however, to 
make security watertight, and security breaches are not uncommon. 
The consequences of stolen credentials go well beyond the leakage of
other types of information because they can further compromise 
other systems.
This paper criticizes the practice of using clear-text identity
attributes, such as Social Security or driver's license numbers---which are in principle not even secret---as acceptable 
authentication tokens or assertions of ownership, and
proposes a simple protocol that straightforwardly applies 
public-key cryptography to make identity claims verifiable, even 
when they are issued remotely via the Internet. This protocol has
the potential of elevating the business practices of credit providers, rental 
agencies, and other service companies that have hitherto exposed consumers 
to the risk of identity theft, to where identity theft becomes virtually impossible.
\end{abstract}

\begin{IEEEkeywords}
Computer Security, Authentication, Identity management systems, Technology social factors, privacy.
\end{IEEEkeywords}

%
\IEEEpeerreviewmaketitle

\section{Introduction}
A recent data breach at consumer credit reporting agency Equifax has compromised 
sensitive personal information such as Social Security numbers, birth dates, 
and driver's license numbers of as many as 143 million US consumers and up to 44 
million British residents~\cite{guardian09}. Consumers are outraged and concerned 
about the looming threat of falling victim to identity theft. Dozens of people 
whose personal
data were leaked have filed lawsuits against Equifax. Politicians decry the 
lack of control that consumers have over their information as it is collected
and sold by companies. Senator Warren and others were the first to respond
with legislation they introduced as the Freedom from Equifax Exploitation (FREE) Act~\cite{warren}, which would let 
consumers ``freeze'' and ``unfreeze'' their accounts at no cost. In the meantime, 
the Federal Trade Commission (FTC) has started an investigation into the hack, 
scrutinizing the company's data security practices for possible neglect.

While all these responses address legitimate concerns regarding the business 
practices of credit reporting agencies and emphasize the necessity 
of companies taking responsibility for the protection of personal 
information that is stored on their servers, one very important question
remains unasked: Why, in this day and age, do we allow plain-text, copyable
strings of characters to serve as a means of authentication?

This paper provides answers to this question, and it suggests a path forward 
by providing a practical solution towards improving the state of affairs. In particular, 
it shows how well-established cryptographic algorithms can be leveraged to 
drastically improve the verifiability of identity claims. The result is a 
secure and verifiable format for the storage and exchange of identity claims,
which, once it establishes itself as an industry standard, would render
plain-text personal information useless for the purpose of committing identity 
fraud. Had this kind of technology been deployed, the Equifax breach would 
have still been problematic from a privacy standpoint, but it would not have 
subjected millions of consumers to a substantial risk of falling victim to 
identity fraud. 

\section{Background}
The problem of identity theft is multifaceted, and it can be studied through many different lenses. While the technical execution of an attack can be explained in terms of computer science, engineering, or even psychology (social engineering~\cite{granger2001social}), the phenomenon also can be treated as a societal problem, and its incidence be explained in the frameworks of law and economics. FTC consumer complaint statistics indicate that identity 
theft has been steadily on the rise throughout the 2000s~\cite{finklea2010identity}. Two very different explanations for this have gained traction in the legal academic literature. On the one hand, Lopucki~\cite{lopucki2002did} argues that the decline of public life and the gradual removal of contact information from public registers has lead to an inability of businesses to authenticate clients. Solove~\cite{solove2002identity}, on the other hand, attributes the problem to a lack of control that consumers have over their personal information. This lack of control increases the likelihood that one's personal information ends up in the hands of criminals.

The thesis of this article is that neither Lopucki nor Solove is right. Of course, impersonation is easier when it can be done over the phone or from behind a computer screen, but publicly available personal information---precisely because it is \emph{publicly} available---can be no useful aid in identifying fraudsters. The lack of control over personal information, however, \emph{is} problematic, but particularly because it is violating of privacy, not because it necessarily weakens security. Besides, given the mega security breaches that we have seen over the recent years (Home Depot, Target, and JPMorgan have had high-profile security breaches, all of which occurred fairly recently, in 2014~\cite{walters2014cyber}), so much data has already been leaked that there is little left to control. The solution to identity theft must therefore not be sought in measures that limit or increase \emph{access} to personal data, but in a mechanism that protects against the improper \emph{usage} of personal data.

\section{Incentives and Risk}
Commercial businesses optimize their activities to maximize net profit. As such, fraud is a risk like any other that needs to be managed. One way to manage risk is to invest in measures that reduce the risk, another is to insure against damages. Companies will make a trade-off between these strategies that they expect is most profitable~\cite{bohme2010security, anderson2006economics}. Preventative measures can reduce insurance expenses and foster consumer trust (which potentially leads to higher revenue), but the cost of implementing preventative measures may outweigh the reduction in insurance cost, in which case the question remains: How pertinent are those measures to the customer experience? If a risk poses no threat to loss of revenue and it is more effective to insure against damages, a company has no incentive to allocate budget for prevention. The only other factors that will tip the scale towards prevention are laws, rules, and regulations which, if not abided by, could hurt business via lawsuits, license revocations, or fines.

Credit card companies are notorious for the poor security of their payment methods: particularly in the US, transactions normally do not require the customer to enter a PIN-code, and in-person transactions typically require no identification; sometimes not even a signature. Purchases made through the Internet often do not involve any authentication at all; it is typically sufficient to enter the card holder name and billing address, along with the credit card number and CCV. It is easy to explain why credit card companies continue to use such primitive technology: it is more profitable to do so. Sophisticated security measures are costly to implement, but, perhaps more importantly, they potentially make it more complicated (or impossible) to make purchases, particularly online, via phone, or in remote areas where direct authentication of a transaction may not be feasible. Hence, improved security, aside from the cost of implementing it, could lead to a reduction in revenue. Apparently, these costs are expected to outweigh the cost of credit card fraud. Instead of improving security to prevent fraud, credit card companies have shifted their focus to just-in-time fraud \emph{detection}, leveraging statistical models (e.g., \cite{4358713}) and machine learning techniques (e.g., \cite{maes2002credit}) to identify suspicious transactions in order to block them or verify their legitimacy with the customer. This turns out to be an adequate solution for the merchant, the consumer, and credit card issuer; the financial damage of credit card fraud is mitigated with minimal impact on the customer experience, and the financial liability is absorbed by the card issuer or the merchant.

As shown in the credit card example, when there exists a direct relationship between a business and its customer, damage control may indeed be a more pragmatic answer to fraud than it is to try to prevent it. The problem is: there might not be such relationship. In the case of Equifax, other businesses, not consumers, are the company's customers. It would be more accurate to say that consumers (or their data) are Equifax's product. Hence, Equifax has little incentive to align its interest with the interests of consumers. On the contrary, safeguarding consumer data is a costly undertaking. If companies like Equifax were more invested in protecting consumer data breaches may be less likely to occur, but impenetrable security is very difficult (if not impossible) to achieve. And when a breach does occur, where a credit card company can simply issue new credit cards and block and revoke compromised ones, data brokers and credit reporting agencies have no such power with respect to the data they handle. This leaves the burden of having to deal with the aftermath of a potential abuse of any stolen information to rest entirely on the affected consumers, even though they never entrusted the company that leaked their information with storing their personal data to begin with.

An example of such an aftermath is the more insidious variant of a credit card fraud that entails the illegitimate creation of a new account, which can take months before it is discovered. Again, it is a misalignment of incentives between businesses and their customers that ends up facilitating identity theft. In \cite{hoofnagle2009internalizing}, Hoofnagle exposes just how embarrassingly easy it is for impostors to obtain credit or medical services. The paper discusses sixteen cases of identity fraud, and in almost all of these cases credit or services were granted on the basis of applications that were rife with errors that should have suggested fraud. Yet, because most of the cost of identity theft is externalized, and businesses make a trade off that optimizes their own profit, they choose to issue credit or provide services even in marginal situations. Companies do not want more rigorous screening because it will cost them revenue. This textbook example of externalizing cost will not disappear unless careless application screening are to become penalized by law, or the injury that results of negligent screening is to be considered a legal basis for a tort claim.

Of course, Equifax is now in a world of legal trouble as it faces lawsuits that are seeking class action status~\cite{ChicagoTribune}. The legal battle that shall unfold in the years to come may lead the industry to re-evaluate its priorities with respect to the protection of consumer data, but that will only address part of the problem. The remainder of this article discusses a technical solution to the data breach problem that, instead of focusing on improvement of data confidentiality, aims to diminish the practical value of plain-text personal information by providing a reliable method for proofs of identity. It is clear, however, that the industry cannot be expected to take the lead in adopting this technology unless a change is incentivised or enforced by law.

\section{Certified Identity Claims}
 The purpose of a Certified Identity Claim (CIC) is for a relying party ({\sf RP}) e.g., an airline, to obtain from a subject ({\sf S}), e.g., a traveler, an attestation of the relationship between {\sf S} and some set of attributes (e.g., last name, date of birth, and passport number)---information similar to that which was compromised by the Equifax breach. The use of CICs is a solution to the problem of identity theft, which entails the fraudulent use of attributes that relate to \emph{another} individual than the person using them, for example, to gain unwarranted access to resources. It is the ability to verify the relationship between an identity claim and the entity that issues it that makes identity theft virtually impossible. \\

\begin{tcolorbox}[title={\sf Sidebar: Encryption and Digital Signatures}, title filled]
Public-key cryptography algorithms use a separate key for encryption and decryption. A key pair consists of a \emph{private} and a \emph{public} key. The private key must be kept secret by its user, while the public key is safe to share with others. Once something is encrypted with a private key, it can only be decrypted with the corresponding public key, and vice versa. 

{\bf Example:} Alice ($A$) and Bob ($B$) have the following key pairs, respectively: $\{A_{pub}, A_{prv}\}$ and $\{B_{pub}, B_{prv}\}$. $A$ and $B$ can now securely exchange messages between each other by encrypting them with each others' public key. A prepares the secret message for B as follows: ${B_{pub}(S)}$. $B$ (and only $B$) can read $S$ by decrypting the message: $B_{prv}(B_{pub}(S))$. The same machinery can be used to issue signatures, simply by swapping the order in which the keys are used. $B$ can apply a signature to a contract $C$ as follows: $B_{prv}(C)$, and anyone, including $A$, can read it using: $B_{pub}(B_{prv}(C))$, and be assured that the message truly originates from $B$. If it were signed with another key than $B_{prv}$, the contents of $C$ would not be readable.
\end{tcolorbox}

The certification of identity attributes is carried out by a trusted third party, referred to as the attribute authority ({\sf AA}). In this system, identity theft would require an attacker to either compromise the security between {\sf S} and {\sf AA} or steal the private key of {\sf AA}. Both are very difficult to achieve, and both are reparable, respectively by resetting {\sf S}'s authentication credentials or by revocation of {\sf AA}'s certificate.

\subsection{Verification of CICs}
\begin{figure}[ht]
    \centering
  \includegraphics[width=.50\textwidth]{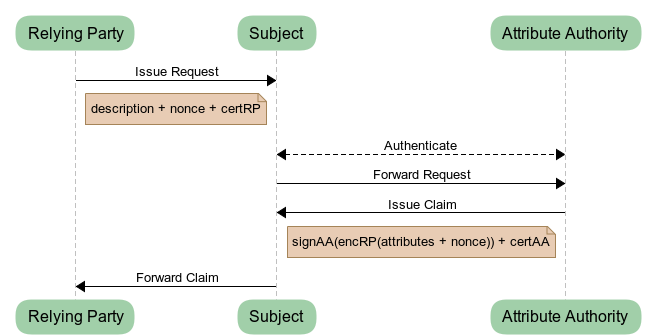}
  \caption{A UML sequence diagram that describes issuance of a CIC.}
  \label{fig:sequence}
\end{figure}

The use of a CIC involves a particular sequence of actions and message exchanges. 
Let us examine the message sequence that is presumably the most common, where {\sf RP} requests {\sf S} to prove ownership of some attribute, and {\sf S} obliges by requesting {\sf AA} to provide a CIC, which {\sf S} finally relays to {\sf RP}. This sequence, illustrated in Fig.~\ref{fig:sequence}, would substitute for requests for \emph{uncertified} identity claims which, on the Web, are ordinarily solicited via a form that is rendered in the browser and filled out manually with keyboard input from the user. An important difference is that while an HTML form provides a description that is understandable by humans, the request sent from {\sf RP} to {\sf S} has to be passed along to a third party, {\sf AA}, which, after authenticating {\sf S}, will have to parse the request and compile an appropriate answer. All of the latter should be carried out automatically, so the format of the request must be machine readable. Importantly, the request must \emph{also} be rendered in a human-readable form to enable the user operating {\sf S} to determine whether it would indeed like to grant the request and forward it to {\sf AA}, or deny it and cancel the exchange.

If we examine Fig.~\ref{fig:sequence} a bit closer, we see that the request consists of three parts; a description, a nonce, and a certificate. The description encodes what particular information is requested from {\sf S}.
The nonce is a randomly generated bit string that is only used once, never across subsequent requests. The nonce is required to prevent replay attacks. Because {\sf AA} bundles the requested attributes with the nonce that is provided by {\sf RP}, it is evident that the certification must have occurred in response to {\sf RP}'s request. Without the nonce, {\sf S} could reuse a CIC that it had previously obtained from {\sf AA}, even if {\sf AA} would no longer certify the claim at the present time (e.g., because {\sf S} no longer has an active account).

After authenticating {\sf S}, {\sf AA} obtains the request and has to formulate a valid response. For instance, if {\sf RP} is a lender and {\sf S} is a potential borrower, the request may look like {\tt [name, credit\_score]}, and the matching attributes would look something like {\tt ['John Davis', 589]}. The next step is to encrypt the attributes together with the nonce using the public key of {\sf RP}. The resulting cipher text is then signed by {\sf AA} (using its private key). The extra round of encryption that is applied, before {\sf AA} does its signing, is critical because 
it prevents {\sf RP} from being able to stage a man-in-the-middle attack. Without it, {\sf RP} would be able to retrieve a request from \emph{another} relying party, say {\sf X}, forward it to {\sf S} (along with the same nonce), obtain a CIC from {\sf S} and submit it to {\sf X}. Neither {\sf X} or {\sf S} would be able to detect the attack. Because {\sf AA} encrypts the attributes with the public key of {\sf RP}, {\sf X} will not be able to read them. It is important that the signature is applied after the encryption and not the other way around. If the keys were to be used in reversed order, {\sf RP} would be able to decrypt the signed attributes and encrypt them with {\sf X}'s public key before forwarding them to {\sf S}. Alternatively, if {\sf RP} were to present {\sf X}'s certificate to {\sf S} instead of its own, the attack would be detectable by {\sf S}.

\subsection{Trust}
Before {\sf RP} can verify a claim, it first has to decide whether {\sf AA} is authoritative to assert the veracity of the claim. The authenticity of the certification provided by {\sf AA} may be derived from the certificate that {\sf RP} receives from {\sf S}. This certificate ties the ``common name'' of the {\sf AA} (e.g., bankofamerica.com) to a public key, attested by a chain of trust that should lead back to an intermediary or root CA that {\sf RP} also trusts. The most trivial way for {\sf RP} to keep track of relationships between {\sf AA}s and the types of claims they are trusted to certify, is to maintain a white list. Simply put, for each claims request that {\sf RP} makes, it will need to \emph{a priori} decide whose certifications it will accept as material. For example, for a claim that establishes holdership of a bank account, all well-established banks could be acceptable {\sf AA}s. Another method would be for {\sf RP} to request a claim from {\sf AA} and expect it to be certified by the American Bankers Association. What this mechanism basically gives rise to is a dynamic variant of a Public-Key Infrastructure (PKI) that is guided by ad-hoc inquisition rather than deliberate top-down one-off design.



\subsection{Privacy}
CICs can be viewed as a privacy-enhancing technology. Apart from preventing misuse of personal information, the use of CICs obviates the need for the collection of personal information beyond what is strictly necessary. Because identity claims are independently verifiable, there is no need to rely on cross referencing with other information in order to gain confidence in the authenticity of the claims. Better yet, a relying party will be less inclined to rely on omni-directional identifiers~\cite{cameron2005laws} (i.e., information that uniquely identifies an entity across multiple contexts, such as a Social Security Number) if it can increase confidence in the veracity of the identity claims it actually cares about. For instance, assume that a service provider needs to determine eligibility on the basis of an applicant's age, taxable income, and marital status. Instead of requesting copies of the applicant's passport, tax return, and marriage certificate, it could simply request CICs that attest the required information, without even revealing as much as the applicant's name.

The decoupling between {\sf RP} and {\sf AA} is deliberate. The fact that all communication between them goes through {\sf S} puts the user in maximal control of the information flow. In this way, the user controls when and how often a third party can access their personal information. Of course, modulo legal provisions in a privacy statement, a user has no control over what a third party does with the personal information it has already acquired. However, particularly for transient claims (e.g., account balance), it is useful that the protocol gives {\sf S} the power to throttle how often {\sf RP} receives updates. Of course, the decisions of {\sf S} may well be performed automatically on the basis of some predefined policy rather than having a human on the loop. The avoidance of direct communication between {\sf RP} and {\sf AA} could potentially also prevent {\sf AA} from tracking the activities of {\sf S}, but the protocol in Fig.~\ref{fig:sequence} reveals {\sf RP}'s identify to {\sf AA} by sharing its public key.

Although a relying party is capable of storing the contents a CIC in clear text, the sequence of cryptographic operations on the information guarantees that {\sf RP} cannot extend the ability to verify the CIC to a third party without also handing over its private key, which would compromise {\sf RP}'s own security. To inspect the CIC, it must first be decrypted with {\sf AA}'s public key and then once again with {\sf RP}'s private key. Without the latter step, the contents of the CIC is unreadable, and if the claim is shared in clear text it is no longer verifiable.

\subsection{Vulnerabilities}
The security of CICs is predicated on some basic assumptions which, albeit reasonable, under certain circumstance could be broken. Let us discuss them one by one.
\subsubsection{Compromised keys}
It is possible that the private key of {\sf RP} or {\sf AA} gets stolen. This would impair the ability of {\sf S} to authenticate them, thus making sessions susceptible to man-in-the-middle attacks. When {\sf AA} has its private key stolen, the trustworthiness of its signature is also compromised.
\subsubsection{Faulty or dishonest parties}
The cryptography prevents attackers from tampering with claims, but it has no way to prevent false information from ending up in a claim. Claims are trusted on the basis of authority. In principle, {\sf AA} could certify a false claim, or it could impersonate {\sf S}. However, this sort of malicious behavior would violate trust and would quickly lead to the demise of {\sf AA}.
\subsubsection{Broken random number generators}
It is important that a sufficiently random nonce is produced by {\sf RP}. If it fails to do so, {\sf S} may be able to successfully ``replay'' a previously issued CIC.
\subsubsection{Quantum computers}
The strength of public-key cryptography hinges on the computational intractability of mathematical problems like prime factorization of large integers. Quantum computers hold the promise of performing factorization in exponentially less time than classical or stochastic methods could~\cite{deutsch1992rapid}. Thus far, quantum computers are not that capable, but there are reasonable expectations that quantum supremacy is achievable in the not-too-distant future~\cite{boixo2016characterizing}.

\subsection{Adoptability}
The proposed protocol can be implemented using off-the-shelf technology and existing web standards (RSA~\cite{rivest1978method} for the signing and verification of claims, X.509~\cite{housley1998internet} for the certificates that authenticate the signatures, HTTPS~\cite{rescorla2000http} for end-point verification and secure communication), and would require only a modest standardization effort. Specifically, the syntax of requests and claims needs to be agreed upon, and there must be a standardized conceptualization of the attributes that can be featured in them: all parties involved will need to understand the contents of a CIC. SAML~\cite{hughes2005security} is an XML-based standard that is already equipped with the ability to encode assertions regarding identity attributes, but the idea of certified identity claims could be extended beyond the assertion of simple attributes. A more elaborate protocol could potentially leverage the W3C Web Ontology Language (OWL)~\cite{bechhofer2009owl} and allow for the formulation of more expressive assertions that involve relationships between multiple entities and certifications by multiple authorities.

An important feature of CICs is that they do not require session-oriented protocols; CICs are self-contained and are tied to a particular request by means of a nonce and a particular {\sf RP} via encryption. This means that the exchange of CICs can be conducted entirely in a RESTful~\cite{fielding2000architectural} fashion. Secure connections are assumed between {\sc S} and {\sc RP} and {\sc S} and {\sc AA}. If these connections are not secure, a man-in-the-middle attack could be staged by an attacker who is interested in letting the victim unwittingly submit claims that are not theirs but the attacker's. For instance, an attacker could hijack an online purchase and reroute the delivery of the purchased item by injecting a false shipping address. An obvious way to secure the message exchange is to use HTTPS, which could also facilitate authentication of {\sc RP} and {\sc AA} with respect to {\sf S}. The methods of authentication of {\sf S} by {\sf RP} and {\sf AA}, respectively, are intentionally left unspecified. Where one organization may want rely on token-based authentication, another may want to use key authentication, a password, biometrics, or some sort of two-factor authentication. We can assume that the level security implemented by {\sf RP} and {\sf AA} is proportional to the sensitivity of the kinds of activities that it protects (e.g., money transfers, access to medical records, etc.). 

\section{Related Work}
Public-key cryptography has been widely adopted as a means to authenticate Web pages (X.509 certificates) and encrypt Web traffic (TLS/SSL). CICs and X.509 certificates are similar, but where a X.509 certificate is designed specifically to tie a public key to a hostname, organization, or individual, a CIC may bear any sort of identity claim. Another difference is that an X.509 certificate is signed by a Certificate Authority (CA), whereas a CIC is signed by an Attribute Authority. This distinction is significant, because the trustworthiness of a CA is determined by its relationship with other CAs, whereas the trustworthiness of an AA is determined with respect to a particular set of attributes, and stands or falls by the relationship that the AA has with the subject.
For example, the non-profit organization Let's Encrypt\footnote{\url{https://letsencrypt.org}} provides free SSL/TLS certificates for any organization or individual that can demonstrate that it controls the domain of the hostname it wishes to tie their public key to. Verification of control is simple and can be automated; the proof requires no more than the creation of a provisioned DNS record and HTTP-accessible resource. But how would just any AA verify whether an individual has any outstanding traffic tickets, for instance? Indeed, it requires an authority with access to very specific information about the subject in order to veritably certify such identity claim. To vouch for the absence of outstanding traffic tickets, the Department of Motor Vehicles would be an bona fide authority, but the American Automobile Association would not.

The concept of claims-based identity was pioneered by Cameron~\cite{cameron2005laws} at Microsoft in the mid-2000s when he laid out seven ``laws'' of identity, which provide useful guidelines for the design of systems that cope with identity on the Internet. Cameron proposes a definition of identity in terms of claims rather than identifiers. With this definition, he shifts the paradigm from identifying individuals to proving the relationship between digital identity and real-world objects. By choosing the term ``claim'' instead of ``assertion,'' he emphasizes that this relationship is always imbued with some level of uncertainty, and could manifest itself as a potential weakness in any secure system. The protocol described in this paper implements the first four of Cameron's laws, which boil down to: ``user consent,'' ``minimal disclosure,'' ``justifiable use,'' and ``context isolation.'' The fifth law states that an identity system needs to allow for pluralism of operators and technologies, which the protocol also does, granted that it leaves the method for end-point authentication unspecified. The sixth and seventh laws are strictly concerned with user experience, which is entirely outside of the scope of this paper.

The notion of \emph{federated} identity captures the fact that identity attributes are often stored across multiple distinct identity management systems. Federated identity management (FIM)~\cite{chadwick2009federated} entails the exchange of identity information between identity management systems. A subset of FIM is concerned with single sign-on (SSO), in which a user's single authentication ticket, or token, is trusted across multiple systems or even organizations. Open standards for FIM are WS-Security and its extension WS-Trust~\cite{anderson2004web}. The goal of FIM is somewhat different from CIC. Where FIM is designed to facilitate long-term collaboration between businesses to provide services to an overlapping customer base, CIC facilitates exchanges between entirely unrelated stakeholders on a per-request basis. Where FIM is aimed at providing tight integration between federations, CIC provides a loose coupling. Popular SSO protocols are OpenID~\cite{recordon2006openid} and OAuth~\cite{hardt2012oauth}. Online platforms like Google\footnote{\url{https://developers.google.com/identity/}} and Facebook\footnote{\url{https://developers.facebook.com/docs/facebook-login/}} also allow third parties to rely on their user authentication mechanism and provide limited access to their users' account details.

Governments have also attempted to design systems for identity management and authentication. Some have become quite successful nationwide identity systems, and are leveraged for purposes as proof of age, proof of citizenship, and for generating digital signatures. These government-run systems, however, tend to poorly interact with commercially-run systems. The National Strategy for Trusted Identities in Cyberspace (NSTIC) that President Obama signed in 2011, sought to change that by creating an ``Identity Ecosystem'' for the improvement of online transactions~\cite{megas2015nstic}. The program went through dozens of pilots, but the idea never materialized in any way, shape, or form that was ready for adoption. Large and complex identity systems come with many risks and challenges~\cite{NAP10346} such as over-centralization, lack of interoperability, possible privacy violations, etc. Basically, the more problems a system attempts to solve, the more challenges it presents to widespread adoption. Highly successful and widely-adopted security mechanisms (e.g., HTTPS) tend to be simple, only address a single issue, and tend not to have too many dependencies on infrastructure.

\section{Conclusions}
The problem of identity theft is a solvable problem; the issue was not prevalent prior
to the advent of Web technology and e-commerce, and its current heavy manifestation 
indicates a problem with the technology we rely on. A lack of sophistication in the way
we verify personal information is to blame. While some suggest that ``data is the 
new currency,''~\cite{gates2014data}, the way we pass around personal information is the 
monetary equivalent of exchanging plain-paper banknotes with handwritten 
denominations on them. This primitive state of affairs is not a reason to scrutinize the security of sites that retain our personal information in order to prevent theft; it is an argument for finding ways to verify the \emph{authenticity} of personal information---similar to how security features on paper money help distinguish genuine bank notes from counterfeit ones. 

The CIC protocol presented in this paper provides a simple solution to the
inexcusable lack of verifiability of identity claims that leads to the tens
of millions of cases of identity theft and tens of billions of dollars of 
fraud damages that are suffered, year after year. The protocol can be 
understood by anyone with basic knowledge of public-key cryptography, it
is inherently distributed, it is fully interoperable with any existing
authentication infrastructure, and it could be seamlessly integrated 
into the online user experience as a substitute of ordinary HTML forms.

Once the use of CICs becomes widespread, the use of clear-text 
identity attributes can be abandoned altogether, which would render 
security breaches, like the one that happened to Equifax, considerably 
less detrimental.
However, it will be up to governments or regulators like the Federal Trade 
Commission to impose restrictions on the acceptance of unverifiable
user-provided electronic data, particularly for the purpose of 
entering legally binding agreements, because most businesses have 
very little incentive to impose such restrictions themselves.

\section*{Acknowledgments}
\addcontentsline{toc}{section}{Acknowledgments}
I would like to thank Prof. Chris Hoofnagle for his encouragement and critical feedback on this work, and for the fruitful discussions I had with him about identity fraud, privacy law, and consumer protections. I am also thankful to Antonio Iannopollo for his helpful suggestions that greatly enhanced the readability of this paper.

\bibliographystyle{abbrv}
\bibliography{Refs}

\end{document}